\documentclass[showpacs,eqsecnum,aps,twocolumn,floatfix]{revtex4}
\usepackage{graphicx,subfigure,amsmath,amssymb}
\usepackage{epsfig}
\usepackage{multirow}
\usepackage{rotating}
\begin{document}

\title{Bohr Hamiltonian with an energy dependent $\gamma$-unstable Coulomb-like potential\footnote{The final publication is available at Springer
via \url{http://dx.doi.org/10.1140/epja/i2016-16314-8}}}

\author{R. Budaca$^{1}$}

\affiliation{$^{1)}$"Horia Hulubei" National Institute for Physics and Nuclear Engineering, Str. Reactorului 30, RO-077125, POB-MG6, M\v{a}gurele-Bucharest, Romania}

\begin{abstract}
An exact analytical solution for the Bohr Hamiltonian with an energy dependent Coulomb-like $\gamma$-unstable potential is presented. Due to the linear energy dependence of the potential's coupling constant, the corresponding spectrum in the asymptotic limit of the slope parameter resembles the spectral structure of the spherical vibrator, however with a different state degeneracy. The parameter free energy spectrum as well as the transition rates for this case are given in closed form and duly compared with those of the harmonic $U(5)$ dynamical symmetry. The model wave functions are found to exhibit properties that can be associated to shape coexistence. A possible experimental realization of the model is found in few medium nuclei with a very low second $0^{+}$ state known to exhibit competing prolate, oblate and spherical shapes.
\end{abstract}
\pacs{21.60.Ev, 21.10.Re, 27.50.+e, 27.60.+j}
\maketitle

\renewcommand{\theequation}{2.\arabic{equation}}
\section{Introduction}
\label{sec:1}
Analytic solutions of the collective Bohr-Mottelson model \cite{Bo1,Bo2} received in the recent years a boost of innovative ideas such as the use of a deformation dependent mass term \cite{M5,M6}, interplay of various shape-phase conditions \cite{Noi1,Noi2,Noi3,Yu}, inclusion of higher order multipole deformations \cite{Bonoct,Guo} as well as of the new solvable potentials \cite{Buganu1,Buganu2,Buganu3,Woods,Rosen,Hulten}. In this trend of alternative approaches, one can also include the adaptation of the formalism based on the energy dependent (non-local) potentials to the quadrupole collective excitations \cite{Eu}. It is interesting that although the variation of the nuclear properties with the energy is a well established fact, the energy dependent potentials are poorly employed in nuclear physics, with only few notable applications regarding quark systems \cite{Lomb1,Sanctis}. On the other hand, some authors expended great effort to ensure the state independence of the potential \cite{Buganu1,Buganu2,Buganu3,Levai1,Levai2,Kharb,Raduta1,Raduta2}.

The energy dependence of the coupling constant of the potential, drastically changes the analytical properties of the associated eigensystem. Therefore, introducing such a concept into the Bohr Hamiltonian must be made with care and with a sufficiently strong phenomenological motivation, because the potential energy depends in general on both shape variables $\beta$ and $\gamma$. The complex collective motion can be exactly separated into a vibrational and rotational components when the potential energy is $\gamma$-independent, due to the fact that the $\gamma$ variable is coupled with the rotational degrees of freedom. In this case of $\gamma$-unstable conditions the energy dependence of the potential is then strictly associated to the vibrational motion. On the other hand, the collective potential defines the nuclear shape, such that its energy dependence imply a shape instability which can be materialized in a shape coexistence \cite{WoodHey}. Shape coexistence is usually marked by the existence of very closely positioned low energy states which correspond to different shapes. It must be understood that it is associated only to extremely distinct shapes such as for example spherical and axially deformed, prolate and oblate, highly separated prolate or oblate deformations. These combinations fit quite well in the $\gamma$-unstable premise of the present approach whose prolate or oblate character is indeterminate.

The Bohr Hamiltonian with a non-local potential was first considered in Ref. \cite{Eu}, in the case of the $\gamma$-unstable five-dimensional harmonic oscillator potential with an increasing string constant with the energy of the system. The simplest energy dependence was used, {\it i.e.} the linear one. Due to mathematical constrictions, the model describes a physical system relevant for collective states only in the asymptotic limit of the slope parameter, where its eigensystem is fully scalable. This provided a new parameter free model called Stiffening Spherical Vibrator (SSV) similar to the harmonic spherical vibrator model $U(5)$ \cite{Bo1} but with distorted spectral properties. SSV enriched thus the set of other parameter-independent collective solutions \cite{Budaca2,Fort,Casten1,Casten2,Cejnar,Buganu} stemmed from the seminal works of Iachello concerning the critical point solutions $E(5)$ \cite{E5} and $X(5)$ \cite{X5}. In this paper, one will refer by $U(5)$ to the spherical vibrator model, even though the general $U(5)$ symmetry contains enough anharmonicities \cite{Castenbook}. The SSV mo\-del but with an energy decreasing string constant leads to an energy spectrum with a threshold corresponding to an infinite quantum number, and therefore does not present any practical importance. However a similar behaviour of the spectrum was obtained in Ref.\cite{FortVit,FortVitu} in connection with a Coulomb-like potential. It would then be interesting to see the effect of a linearly energy dependent coupling constant on the results corresponding to a $\gamma$-unstable Coulomb-like potential.

In this paper one will show that the situation is quite reversed in respect to the formalism of Ref.\cite{Eu}, meaning that the unperturbed spectrum of the $\gamma$-unstable Coulomb-like potential is expanded, reaching in the asymptotic limit of the slope parameter the vibrational energy level sequence. The use of a singular potential instead of a confining one leads however to quite different degeneracies of these energy levels. This aspect provides the opportunity to theoretically interpret some very low lying $0^{+}$ states through such complex $\beta$ excitations. These states are considered as a signature for the shape coexistence phenomenon and their emergence is usually ascribed to the proximity of the corresponding nuclei to major shell and subshell closures where there is a heightened interplay between single-particle and collective degrees of freedom. Although the theoretical description of the phenomenon is usually approached with deformed mean-field models or shell-model calculations using different effective interactions extensively reviewed in Ref. \cite{HeyWood} and more recently with interacting boson model based approaches \cite{Mc,Thomas1,Thomas2,Garcia1,Garcia2,Garcia3,Zhang,Nomura}, the interpretation of the resulting low lying energy spectrum is often made with the aid of a collective wave-function or by mappings in collective coordinates. Therefore, a fully collective description of these $0^{+}$ states is more than justified.

The proposed scope is achieved first by presenting in the next Section the analytical construction of the model regarding the Hamiltonian, its general solutions, electromagnetic properties and the characteristics of the aforementioned asymptotic limit. The associated numerical a\-na\-lysis of the model is given in Section III by means of various numerical applications mostly aimed at the comparison of the asymptotic limit with the $U(5)$ model and few experimental data of nuclei considered as its candidates. Finally, the conclusions and some perspectives of the proposed theoretical formalism are presented in the last Section.

\renewcommand{\theequation}{2.\arabic{equation}}
\setcounter{equation}{0}
\section{The theoretical framework}
\label{sec:2}
\subsection{The model Hamiltonian}
\label{subsec:1}

The general Bohr Hamiltonian reads:
\begin{eqnarray}
H&=&-\frac{\hbar^{2}}{2B}\left[\frac{1}{\beta^{4}}\frac{\partial}{\partial{\beta}}\beta^{4}\frac{\partial}{\partial{\beta}}+\frac{1}{\beta^{2}\sin{3\gamma}}\frac{\partial}{\partial\gamma}\sin{3\gamma}\frac{\partial}{\partial\gamma}\right.\nonumber\\
&&\left.-\frac{1}{4\beta^{2}}\sum_{k=1}^{3}\frac{Q_{k}^{2}}{\sin^{2}{\left(\gamma-\frac{2}{3}\pi k\right)}}\right]+V(\beta,\gamma).
\label{BM}
\end{eqnarray}
where by $Q_{k}(k=1,2,3)$ are denoted the operators of the total angular momentum projections on the axes of the intrinsic reference frame, while $B$ is the mass parameter. $\gamma$-instability condition means that the potential is independent of the $\gamma$ shape variable, {\it i.e.} $V(\beta,\gamma)=V(\beta)$. Following the usual steps in case of $\gamma$-unstable solutions, the corresponding eigenvalue equation is separated by factorizing the total wave function as $\psi(\beta,\gamma,\Omega)=F(\beta)\Phi(\gamma,\Omega)$ into a $\beta$ part:
\begin{equation}
\left[-\frac{1}{\beta^{4}}\frac{\partial}{\partial{\beta}}\beta^{4}\frac{\partial}{\partial{\beta}}+\frac{\Lambda}{\beta^{2}}+v(\beta)\right]F(\beta)=\epsilon F(\beta),
\label{b}
\end{equation}
where $\epsilon=(2B/\hbar^{2})E$ and $v=(2B/\hbar^{2})V$ are the reduced energy and potential, and a $\gamma$-angular one:
\begin{eqnarray}
\left[-\frac{1}{\sin{3\gamma}}\frac{\partial}{\partial\gamma}\sin{3\gamma}\frac{\partial}{\partial\gamma}+\frac{1}{4}\sum_{k=1}^{3}\frac{Q_{k}^{2}}{\sin^{2}{\left(\gamma-\frac{2}{3}\pi k\right)}}\right]\nonumber\\
\times\Phi(\gamma,\Omega)=\Lambda\Phi(\gamma,\Omega).
\end{eqnarray}
The last equation corresponding to $\gamma$-angular coordinates was solved by B\`{e}s \cite{Bes} with the following result for the separation constant
\begin{equation}
\Lambda=\tau(\tau+3),
\end{equation}
$\tau$ being the seniority quantum number \cite{Rakavy} associated to the eigenvalue of the second order $SO(5)$ Casimir operator. For each $\tau$ there are multiple realizations of the angular momentum $L$ and its projection $K$ on the intrinsic $z$ axis. The algorithm to determine branching of representations is thoroughly explained in Ref.\cite{Deb} and can be summarized as:
\begin{eqnarray}
\tau=K+3\nu_{\Delta},\,\,\,\nu_{\Delta}=0,1,2,...[\tau/3],
\end{eqnarray}
where $\nu_{\Delta}$ is the missing quantum number in the $SO(5)\supset SO(3)$ group reduction, while square brackets mean the integer part. Finally, the angular momentum takes all integer values between $K$ and $2K$ excluding the $2K-1$ value.

Using the change of function $f(\beta)=\beta^{2}F(\beta)$, the $\beta$ part equation can be written in a canonical-like form
\begin{equation}
\left[-\frac{\partial^{2}}{\partial{\beta^{2}}}+\frac{\left(\tau+\frac{3}{2}\right)^{2}}{\beta^{2}}-\frac{1}{4\beta^{2}}+v(\beta)\right]f(\beta)=\epsilon f(\beta),
\label{difb}
\end{equation}
which is suitable if the $\beta$ potential is chosen to be of the Coulomb type \cite{FortVit}:
\begin{equation}
v(\beta)=-\frac{A}{\beta},\,\,\,A>0.
\label{pot}
\end{equation}
In this paper however, one considers a coupling constant which depends linearly on the energy of the system
\begin{equation}
A\,\,\rightarrow\,\,A(\epsilon),\,\,\,A(\epsilon)=1+a\epsilon.
\end{equation}
In the case of energy dependent potentials, the definition of the density probability or the scalar product must be modified in order to satisfy the continuity equation \cite{Lepage,Saz,Form}. Therefore, the new $\beta$ density probability for a state $\{i\}$ is defined as
\begin{equation}
\rho_{i}(\beta)=\left|F_{i}(\beta)\right|^{2}\left[1-\frac{\partial{v(\epsilon_{i})}}{\partial{\epsilon_{i}}}\right]=\left|F_{i}(\beta)\right|^{2}\left(1+\frac{a}{\beta}\right),
\end{equation}
and all scalar products involving functions of $\beta$ must be amended with the same factor. Consequently, the slope parameter $a$ must be positive given the repulsive nature of the potential (\ref{pot}) and the condition that the density probability to be positive definite in order to describe a physical system. Moreover, the use of linear dependence on energy has the advantage of producing energy independent integration measure for the scalar products. This is actually one of the reasons why this particular energy dependence is usually considered in literature \cite{Lomb1,Lepage,Saz,Form,Yekken1,Yekken2,Garcia}.

\subsection{Solutions}

The procedure for solving the associated Schr\"{o}dinger equation with the energy dependent potential (\ref{pot}), is the same as in the case of the state independent Coulomb potential \cite{FortVit}. Basically, the differential equation (\ref{difb}) is brought to a Whittaker form \cite{Erdely}:
\begin{equation}
\left[\frac{\partial^{2}}{\partial{x^{2}}}-\frac{1}{4}+\frac{k}{x}+\frac{\left(\frac{1}{4}-\mu^{2}\right)}{x^{2}}\right]f(x)=0,
\end{equation}
by the change of variable $x=2\sqrt{\varepsilon}\beta$ together with the following notations:
\begin{equation}
\varepsilon=-\epsilon,\,\,\,k=\frac{A(-\varepsilon)}{2\sqrt{\varepsilon}},\,\,\mu=\tau+\frac{3}{2}.
\end{equation}
The solutions of this differential equation are known as Whittaker functions $M_{\mu,k}$ \cite{Whitt} and can be written in terms of hypergeometric functions of first kind ${}_{1}F_{1}(b,c;x)$ \cite{Abram}, such that:
\begin{equation}
M_{k,\mu}(x)=x^{\mu+\frac{1}{2}}e^{-\frac{x}{2}} {}_{1}F_{1}\left(\mu+\frac{1}{2}-k,2\mu+1;x\right).
\end{equation}
Although, the condition of negative energy guarantees the regularity of the above solution in the origin, the function in general diverges in the asymptotic limit of $x$. This obstacle is circumvented if the hypergeometric function becomes an associated Laguerre polynomial, {\it i.e.} when the first argument is a negative integer:
\begin{equation}
\mu+\frac{1}{2}-k=\tau+2-\frac{A(-\varepsilon)}{2\sqrt{\varepsilon}}=-n.
\label{M}
\end{equation}
This condition together with $\varepsilon=-\epsilon$ provides us with a quadratic equation for the reduced energy:
\begin{equation}
\epsilon=-\frac{(1+a\epsilon)^{2}}{4(n+\tau+2)^{2}},
\label{eqe}
\end{equation}
whose two solutions are
\begin{eqnarray}
\epsilon_{n\tau}^{\pm}&=&\frac{1}{a^{2}}\left[-2(n+\tau+2)^{2}-a\right.\nonumber\\
&&\left.\pm2(n+\tau+2)\sqrt{(n+\tau+2)^{2}+a}\right].
\label{esol}
\end{eqnarray}
The corresponding total $\beta$ wave function is then defined as:
\begin{equation}
F_{n,\tau}(\beta)=\mathcal{N}_{n,\tau}\beta^{\tau}e^{-\eta_{n,\tau}\beta}L_{n}^{2\tau+3}(2\eta_{n,\tau}\beta),
\label{fb}
\end{equation}
where
\begin{equation}
\eta_{n,\tau}=\frac{1+a\epsilon_{n,\tau}}{2(n+\tau+2)},
\end{equation}
while $\mathcal{N}_{n,\tau}$ is the normalization constant determined from the condition
\begin{equation}
\int_{0}^{\infty}\left[F_{n,\tau}(\beta)\right]^{2}\beta^{4}\left(1+\frac{a}{\beta}\right)d\beta=1.
\end{equation}
Using the properties of the associated Laguerre polynomials, one can readily obtain its analytical expression:
\begin{equation}
\mathcal{N}_{n,\tau}=\left(2\eta_{n,\tau}\right)^{\tau+2}\sqrt{\frac{\eta_{n,\tau}n!}{(n+2\tau+3)!(a\eta_{n,\tau}+n+\tau+2)}}.
\label{norm}
\end{equation}
In order to have $\eta_{n,\tau}>0$ and consequently a factor $e^{-\eta_{n,\tau}\beta}$ from the wave function (\ref{fb}) which does not diverge at $\beta\to\infty$, one must choose the "+" sign in the energy expression (\ref{esol}). Further on, one will drop the notation $\pm$, retaining only the solution with the plus sign. Note that although the energy spectrum (\ref{esol}) depends on a single quantum number defined by the sum $n+\tau$, the corresponding eigenfunctions have a separate dependence on $n$ and $\tau$, respectively.

\subsection{$E2$ electromagnetic transitions}

Employing the general expression for the quadrupole transition operator,
\begin{eqnarray}
T_{\mu}^{(E2)}&=&t\beta Q_{\mu},\\
Q_{\mu}&=&D_{\mu0}^{2}(\Omega)\cos{\gamma}+\frac{1}{\sqrt{2}}\left[D_{\mu2}^{2}(\Omega)+D_{\mu-2}^{2}(\Omega)\right]\sin{\gamma}\nonumber,
\label{TE2}
\end{eqnarray}
where $t$ is a scaling factor, one can calculate the transition rates using the wave function (\ref{fb}) derived above. The final result for the $E2$ transition probability is given in a factorized form:
\begin{eqnarray}
&&B(E2;n\tau L\rightarrow n'\tau'L')=t^{2}(\tau',L';1,2||\tau,L)^{2}\nonumber\\
&&\times\left[\langle\tau|||Q|||\tau'\rangle B_{n\tau;n'\tau'}\right]^{2},
\label{be2}
\end{eqnarray}
where $(\tau_{1},L_{1};\tau_{2},L_{2}||\tau_{3},L_{3})$ is the $SO(5)$ Clebsch-Gordan coefficient dictating the angular momentum selection rules with the multiplicity entry omitted due to the fact that the relevant states, {\it i.e.} those with small $\tau$ have the multiplicity 1 \cite{Rowebook}. The usually encountered values of these coefficients are tabulated in Ref.\cite{RoTu} where a general calculation prescription is also presented. The corresponding non vanishing reduced matrix element has a simple form in respect to the seniority $\tau$ \cite{Rowe1,Rowe2}:
\begin{equation}
\langle\tau|||Q|||\tau'\rangle=\sqrt{\frac{\tau}{2\tau+3}}\delta_{\tau,\tau'+1}+\sqrt{\frac{\tau+3}{2\tau+3}}\delta_{\tau,\tau'-1},
\end{equation}
while $B$ is the integral over the $\beta$ shape variable with the modified integration measure:
\begin{equation}
B_{n\tau;n'\tau'}=\int_{0}^{\infty}F_{n,\tau}(\beta)F_{n',\tau'}(\beta)\beta^{5}\left(1+\frac{a}{\beta}\right)d\beta.
\label{bint}
\end{equation}
This integral can be brought to a closed analytical form by using the properties of the associated Laguerre polynomials explained in the Appendix.

\subsection{The asymptotic limit}

From the dependence of the energy function (\ref{esol}) on the slope parameter $a$ for different quantum numbers $N=n+\tau$, one can observe that the whole spectrum presents a convergent behaviour at very high values of $a$. As a matter of fact, in the asymptotic limit of $a$ the energy attains the expression:
\begin{equation}
\epsilon_{n\tau}^{(asymp)}=-\frac{1}{a}+\frac{2}{a^{3/2}}(N+2),
\label{eas}
\end{equation}
which when normalized to the ground state and divided to the excitation energy of the first excited state provides the same energy level sequence as the five-dimensional harmonic oscillator model \cite{Bo1,Bo2}. As a consequence, the slope parameter $a$ becomes a simple scaling factor and one obtains another parameter-free collective solution which hereafter will be called Asymptotic Energy Dependent Coulomb (AEDC) model. The essential difference from the $U(5)$ spectrum is the quantum number assignment. Indeed, the $U(5)$ states are indexed by the quantum number $N_{h.o}=2n+\tau$, whereas in the present case by $N=n+\tau$.

The parameter-free character of the model in the a\-symptotic limit is also reflected on the wave functions. Namely, the energy dependence of the wave function vanishes as
\begin{equation}
\eta_{n,\tau}^{(asymp)}=\frac{1}{\sqrt{a}}.
\label{lim}
\end{equation}
Such that the asymptotic $\beta$ wave function can be expressed as:
\begin{equation}
F_{n,\tau}^{(asymp)}(\beta)=\mathcal{N}_{n,\tau}^{(asymp)}\beta^{\tau-\frac{1}{2}}e^{-\frac{\beta}{\sqrt{a}}}L_{n}^{2\tau+3}\left(2\beta/\sqrt{a}\right),
\label{fbas}
\end{equation}
with
\begin{equation}
\mathcal{N}_{n,\tau}^{(asymp)}=\left(\frac{2}{\sqrt{a}}\right)^{\tau+2}\sqrt{\frac{n!}{(n+2\tau+3)!}}.
\end{equation}
The normalisation constant is considered here in respect to the usual $\beta^{4}d\beta$ integration measure, because the dominant correction term $a/\beta$ of the density probability is now included in the asymptotic wave function. It is worth to mention that its expression can also be derived from (\ref{norm}) in virtue of (\ref{lim}). The slope parameter $a$ has a clear scaling role, such that the asymptotic expression of the integral (\ref{bint}) implied in the calculation of $E2$ transition probabilities:
\begin{equation}
B_{n\tau;n'\tau'}^{(asymp)}=\int_{0}^{\infty}F^{(asymp)}_{n,\tau}(\beta)F^{(asymp)}_{n',\tau'}(\beta)\beta^{5}d\beta,
\end{equation}
when normalized will lose any dependence on $a$. Moreover, the composition of the $\beta$ wave function (\ref{fbas}) is somewhat similar to that from the five-dimensional harmonic oscillator model:
\begin{equation}
F_{n,\tau}^{(h.o.)}(\beta)=\sqrt{\frac{2n!}{\Gamma\left[n+\tau+\frac{5}{2}\right]}}\beta^{\tau}e^{-\frac{\beta^{2}}{2}}L_{n}^{\tau+\frac{3}{2}}\left(\beta^{2}\right).
\label{fu5}
\end{equation}
This is not surprising given the identical energy level scheme of the two models.

Finally, plugging expression (\ref{eas}) into the energy dependent potential, one obtains:
\begin{equation}
v^{(asymp)}(\beta)=-\frac{2(N+2)}{\sqrt{a}\beta}.
\end{equation}

\renewcommand{\theequation}{3.\arabic{equation}}
\section{Numerical results}
\label{sec:3}

\begin{figure}[t!]
\begin{center}
\includegraphics[width=0.45\textwidth]{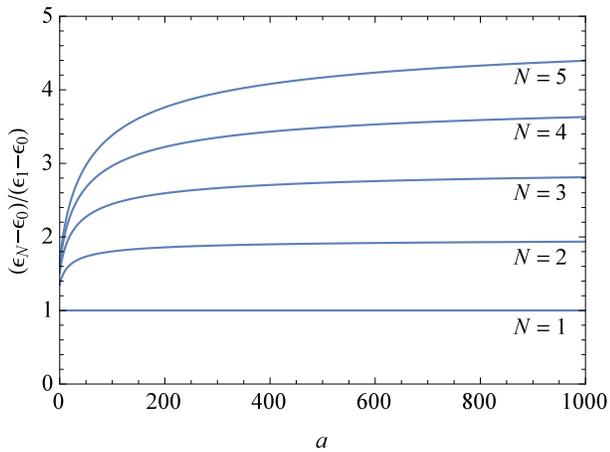}
\end{center}
\caption{The energy spectrum given by Eq.(\ref{esol}) normalized to the energy of the first excited state and with ground state energy fixed to zero is given as function of the slope parameter $a$. The curves are indexed by $N=n+\tau$.}
\label{aspec}
\end{figure}

In the asymptotic regime of the parameter $a$, the ratio $R_{4/2}$ of the excitation energies corresponding to the first two excited states reaches the well known collective minimum value 2 associated to the two-phonon state of the five-dimensional spherical vibrator model. Therefore, only this situation has a practical use. Nevertheless, in order to understand the analytical properties of the asymptotic regime, an overall inspection of the influence of the slope parameter $a$ is necessarily required. The dependence of the normalized energy spectrum defined by Eq.(\ref{esol}) on the slope parameter $a$ is plotted in Fig. \ref{aspec} for few values of the global quantum number $N=n+\tau$. The saturation of the spectrum at high $a$ is obvious, however the vibrational-like energy sequence is achieved at much higher values and with an increasing convergence radius. For example, in order to achieve an accuracy of $10^{-3}$ for the first 3 excited states, one must imply for $a$ a value of order $10^{6}$. From the same figure one can also see that when $a\to0$ one recovers the usual $\gamma$-unstable model with a local Coulomb-like $\beta$ potential \cite{FortVit}, which is not obvious from the energy expression (\ref{esol}). The zero limit can be recovered by factorizing the energy (\ref{esol}) with $(n+\tau+2)^{2}/a^{2}$ which is then approximated by $1/a^{2}$.

\begin{figure*}[t!]
\begin{center}
\includegraphics[width=0.9\textwidth]{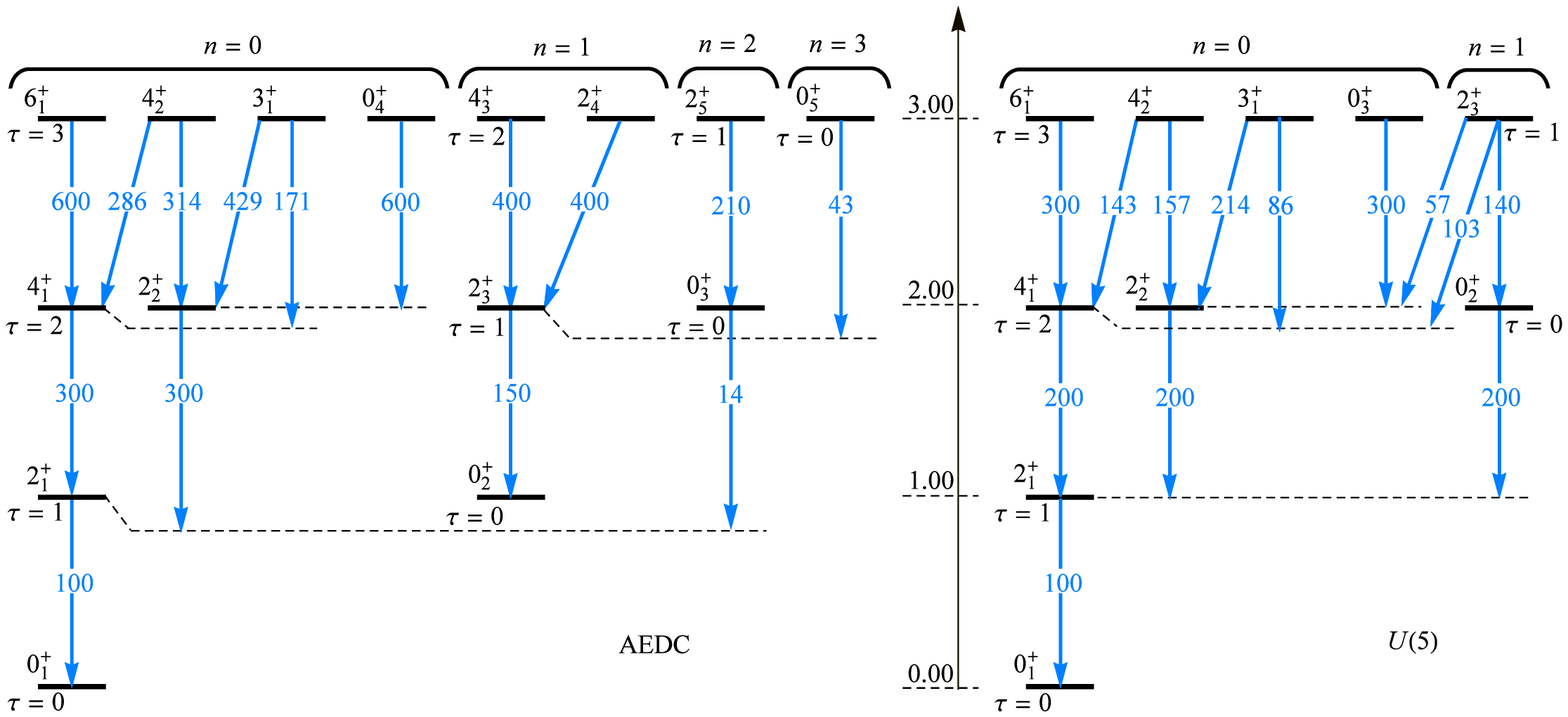}
\end{center}
\caption{Predictions of the present model (AEDC) for the lowest portion of the spectrum are compared with the corresponding level scheme of the spherical vibrator model. The spectrum is normalized to the ground state energy and given in units of the first excited state energy. Similarly, the $E2$ transition rates are given in terms of $B(E2,2^{+}_{1}\rightarrow0^{+}_{1}) =100$. The states in both cases are indexed by order in energy scale with the quantum number assignment also schematically indicated.}
\label{comp}
\end{figure*}

In what follows one will concentrate on the numerical applications regarding the asymptotic regime of the model. As its similarity to the $U(5)$ energy level scheme was already mentioned, a close comparison between the two models is compulsory. Fig. \ref{comp}, where the low lying energy spectra and the corresponding quadrupole electromagnetic transition probabilities normalized to the same quantities corresponding to the first excited ground band state are depicted for both models, serves perfectly this purpose. For the calculus of the $E2$ transition rates defined through Eq.(\ref{be2}) one gathered the values for the $SO(5)$ Clebsch-Gordan coefficients from Ref.\cite{Rowebook,Rowe1}, and used the result from the Appendix for computing the relevant $\beta$ integrals. The $U(5)$ transition rates are taken from \cite{Deb}. Due to the common $SO(5)$ upbringing, the distribution of states by seniority is identical in both cases. The difference appears in the $\beta$ excited states, which are shifted down in the present model in respect to the spherical vibrator spectrum. This also induces a different degeneracy of the energy levels reflected in a higher density of energy degenerate states. In what concerns the $E2$ transition rates, the $\Delta n=0$ ones are higher in the present model retaining however a similar trend of relative distribution of values as in the $U(5)$ case. While the interband transitions are completely different in the relating states and therefore cannot be compared. Indeed, in the present model the $\Delta n=1$ transitions which were reported as very small for the unperturbed $\gamma$-unstable Coulomb-like potential \cite{FortVit}, vanish altogether. The only nonvanishing interband transitions are those with $\Delta n=2$, which are quite few in this region and whose values are still small but comparable with the reference value of $B(E2,2^{+}_{1}\rightarrow0^{+}_{1})$. This points to the fact that consecutive $\beta$ excited bands are fully decoupled and the only allowed interband transitions are through two vibrational quanta.

\begin{figure}
\begin{center}
\includegraphics[width=0.45\textwidth]{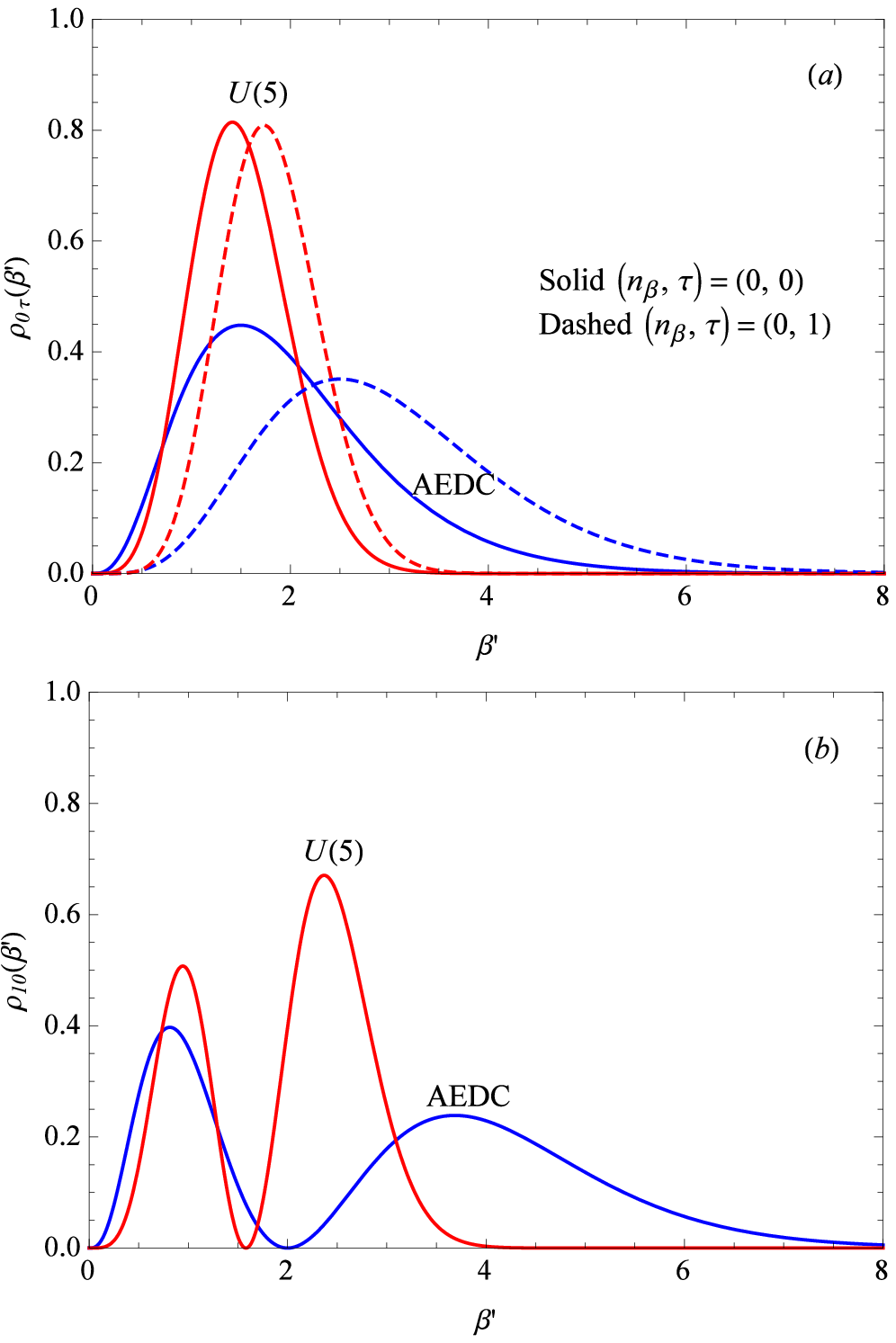}
\end{center}
\caption{(a) Ground state and first $\tau$ excited state $\beta$ density probability in the present and $U(5)$ cases as function of $\beta'$ which is just $\beta$ for $U(5)$ model and respectively $\beta'=\beta/\sqrt{a}$ for AEDC. (b) The same but for the first $\beta$ excited state density probability.}
\label{ro}
\end{figure}

As a consequence, the similarity with the energy level sequence of the spherical vibrator model seems to be a fortunate coincidence, because the proposed model has quite distinct analytical properties. Indeed, even though the corresponding $\beta$ wave functions (\ref{fbas}) and (\ref{fu5}) have the same factorized expression, the distinct role played by the seniority and the $\beta$ vibrational quantum numbers induce major differences in the $\beta$ probability density distribution contrary to the energy spectrum which depends only on $N=n+\tau$. This can be clearly seen from Fig. \ref{ro}, where one plotted the density probability distribution
\begin{equation}
\rho_{n,\tau}^{(i)}(\beta')=\left|F_{n,\tau}^{(i)}(\beta')\right|^{2}\beta'^{4},
\end{equation}
for the ground state, first $\gamma$-angular and $\beta$ excited states in both AEDC and $U(5)$ cases. Here $i$ stands for "$h.o.$" or "$asymp$", and $\beta'$ is a scaled shape variable which is just $\beta$ for the $U(5)$ case, while in the present model's situation is given by $\beta'=\beta/\sqrt{a}$. The peak of the ground state density probability distribution in the $U(5)$ case is sharp and symmetrical, while in AEDC model, it is asymmetrical and extended over a larger interval of $\beta'$ values. This remains true also for the excited states from the ground band. Additionally, the peak of the density distribution shifts and decreases quicker for the AEDC when considering ground band excited states. In what concerns the first $\beta$ excited state, it is well known that the associated density probability is split into two peaks. While the $U(5)$ highest probability $\beta$ value from the ground state has an almost zero probability to occur in the first beta excited state, in the AEDC case this possibility is not negligible due to the still maintained asymmetric shape of the peaks. However the most striking contradistinction between the two models comes from the opposite relative height of the two peaks which are also very separated for the AEDC probability. This is consistent with the fact that the main difference between the energy spectra of both models comes from the distribution of the $\beta$ excited states. Taking as a reference the behaviour of the $\beta$ excited density distribution, one can conclude that the phenomenological conditions associated to the present model describe a highly anharmonic oscillation of the nuclear surface. A similar analysis of the quantum fluctuations in the collective $0^{+}$ states was performed in \cite{Chen} for deformed nuclei. As a matter of fact, anharmonicities play an important role in describing the nuclear spectra of near-spherical nuclei \cite{Budaca2,Budaca1}. However, one must note here that regardless of the energy dependence, in the present formalism one used a singular potential and obtained an energy level scheme similar to a collective model with a confined potential, {\it i.e.} harmonic oscillator.

\begin{figure*}[t!]
\begin{center}
\includegraphics[width=0.45\textwidth]{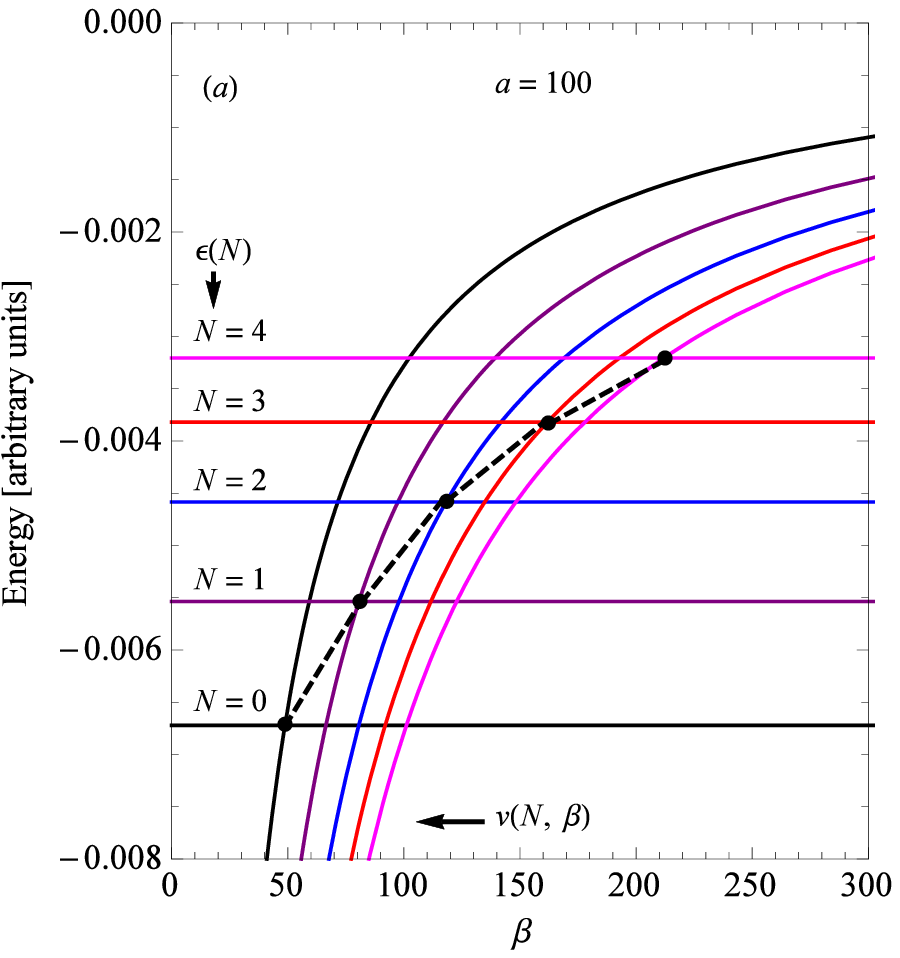}
\includegraphics[width=0.45\textwidth]{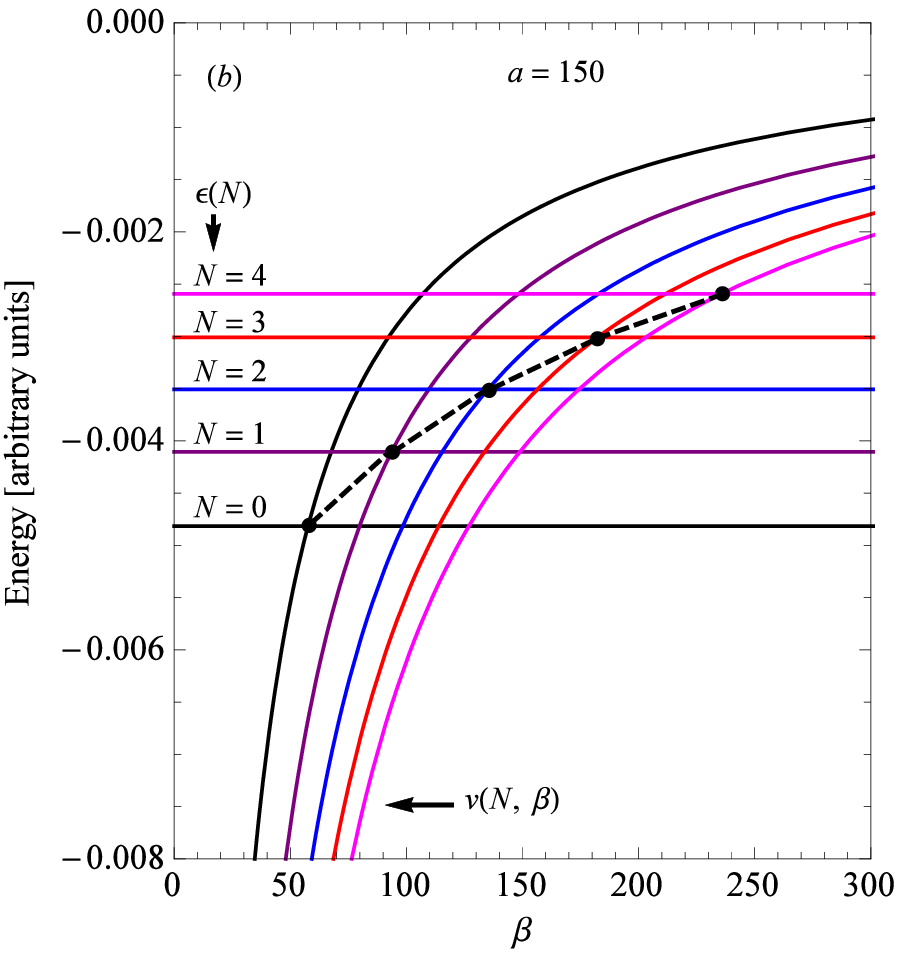}
\end{center}
\caption{Energy levels for first few states indexed by quantum number $N=n+\tau$ are visualized together with their associated state dependent potentials $v(\varepsilon(N),\beta)=v(N,\beta)$ for $a=100$ (a) and $a=150$ (b). For clarity, the corresponding intersections are marked with expanded dots which are linked by straight lines in order to simulate a smooth evolution.}
\label{epot}
\end{figure*}

\begin{table*}
\caption{Experimental low lying energy spectra of $^{72}$Se \cite{72sekr}, $^{74}$Se \cite{74se}, $^{72}$Kr \cite{72sekr}, $^{98}$Mo \cite{98mo} and $^{100}$Mo \cite{100mo} are compared with the theoretical results from the asymptotic limit. Values in parentheses denote states with uncertain assignment of angular momentum.}
\label{tabe}
\begin{center}
\begin{tabular}{ccccccc}
\hline\noalign{\smallskip}
~~$L_{n,\tau}$~~&~~Th.~~&~~$^{72}$Se~~&~~$^{74}$Se~~&~~$^{72}$Kr~~&~~$^{98}$Mo~~&~~$^{100}$Mo~~\\
\noalign{\smallskip}\hline\noalign{\smallskip}
$2_{0,1}$ &1.000& 1.000 & 1.000 & 1.000& 1.000 & 1.000\\
$0_{1,0}$ &     & 1.087 & 1.345 & 0.945& 0.933 & 1.298\\
\noalign{\smallskip}\hline\noalign{\smallskip}
$4_{0,2}$ &2.000& 1.899 & 2.148 &1.862& 1.918  & 2.121\\
$2_{0,2}$ &     & 1.527 & 1.999 &     & 1.819  & 1.986\\
$2_{1,1}$ &     & 2.319 &       &     & 2.233  & 2.733\\
$0_{2,0}$ &     &       &(2.611)&     & 2.493  &\\
\noalign{\smallskip}\hline\noalign{\smallskip}
$6_{0,3}$ &3.000& 2.861 & 3.516 & 2.977&(2.976)& 3.449\\
$4_{0,3}$ &     &       & 3.321 &      & 2.824 & 3.308\\
$3_{0,3}$ &     &(3.000)& 2.969 &      & 2.673 & 3.001\\
$0_{0,3}$ &     &       &       &      & 2.805 & 2.809\\
$4_{1,2}$ &     &       &       &      &(2.964)&\\
$2_{1,2}$ &     &(2.494)&(2.897)&      & 2.802 & \\
$2_{2,1}$ &     &(2.661)&(3.646)&      & 2.964 & 3.298\\
$0_{3,0}$ &     &       &       &      &       & 3.804\\
\noalign{\smallskip}\hline\noalign{\smallskip}
$8_{0,4}$ &4.000& 3.973 & 5.039 & 4.380& 4.155 & 4.906\\
$10_{0,5}$&5.000& 5.225 &       & 6.049&       & 6.286\\
$12_{0,6}$&6.000& 6.623 &       & 7.959&       & \\
$14_{0,7}$&7.000& 8.164 &       &10.085&       & \\
\noalign{\smallskip}\hline
\end{tabular}
\end{center}
\end{table*}

\begin{table*}
\caption{Several commonly available experimental $E2$ transition probabilities for $^{72}$Se \cite{72sekr}, $^{74}$Se \cite{74se}, $^{72}$Kr \cite{krtr} $^{98}$Mo \cite{98mo} and $^{100}$Mo \cite{100mo} are confronted with the model's predictions. All transition rates are normalized to the $2_{1}^{+}\rightarrow0^{+}_{1}$ transition which is set to 100\%.}
\label{tabt}
\begin{center}
\begin{tabular}{crrrrrr}
\hline\noalign{\smallskip}
$L_{n,\tau}\rightarrow L'_{n',\tau'}$&~~Th.~~&~~~$^{72}$Se~~~&~~~$^{74}$Se~~~&~~~$^{72}$Kr~~~&~~~$^{98}$Mo~~~&~~~$^{100}$Mo~~~\\
\noalign{\smallskip}\hline\noalign{\smallskip}
$4_{0,2}\rightarrow2_{0,1}$ &300& 232(27) & 190(10) & 336(92) &214(6)    & 186(11)\\
$2_{0,2}\rightarrow2_{0,1}$ &300&         & 114(33) &         &162(20)   & 138(14)\\
$2_{1,1}\rightarrow0_{1,0}$ &150&         &         &         &36(2)     & 38(11)\\
$2_{1,1}\rightarrow0_{0,0}$ &~~0&         &         &         &$\approx0$& 1\\
$6_{0,3}\rightarrow4_{0,2}$ &600& 274(29) & 171(36) &         &50(2)     & 254(38)\\
$4_{0,3}\rightarrow4_{0,2}$ &286&         & 57(21)  &         &          & 77(16)\\
$4_{0,3}\rightarrow2_{0,2}$ &314&         & $<$40   &         &          & 81(16)\\
$3_{0,3}\rightarrow4_{0,2}$ &171&         & 24(19)  &         &          &\\
\noalign{\smallskip}\hline
\end{tabular}
\end{center}
\end{table*}

For model's candidates one searched nuclei with a vib\-ra\-ti\-o\-nal-like spectrum, but for which the first $2^{+}$ and the second $0^{+}$ states are degenerate. As a matter of fact, the presence of the low lying $0^{+}$ state can be considered as the smoking gun of the present theoretical model which also is the fingerprint of shape coexistence. Such candidates were found in few lighter nuclei from the region which marks the emergence of collective excitations. These are the $^{72}$Se, $^{74}$Se, $^{72}$Kr nuclei placed below the neutron $N=50$ shell closure and the $^{98}$Mo and $^{100}$Mo nuclei positioned above it. All considered nuclei are well known as typical examples of shape coexistence. Indeed, their potential energy surfaces calculated with state-of-the-art energy density functionals are flattened and with multiple minima extended between oblate and prolate shapes \cite{Hilaire}. It is worth to mention here that there is not necessarily a one to one relation between the microscopic potential energy surface and the potential used in the Bohr model. The latter is dictated by a sufficiently fitting description of experimental data corroborated with a phenomenological motivation for the behaviour of the density probability provided by its corresponding wave functions. As a matter of fact, the $\beta$ density distribution depicted in Fig.\ref{ro} shows features compatible with the shape coexistence phenomenon. For example, the non symmetrical and extended profile of the ground state distribution suggests multiple ground state deformations which in view of the $\gamma$-unstable character of the model can be ascribed either to a prolate or oblate shape. Indeed, the microscopic calculations predict quite close absolute values for the coexisting prolate and oblate deformations in the ground states for some of the considered nuclei \cite{Hilaire,Bender,Hinohara1,Hinohara2}. In what concerns the density probability of the first $0^{+}$ excited state, it is consistent with the microscopic findings of Ref.\cite{Hinohara1} regarding the $^{70,72}$Se isotopes. The two peaks of the $0^{+}$ $\beta$ density probability from Fig.\ref{ro}(b) correspond to the two turning points of the surface oscillation, with a clear preference for the smallest deformation which maintain a harmonic behaviour as in the spherical case. In contradistinction, the higher deformation peak is very extended and far from the equilibrium deformation $\rho_{10}(\beta=2)=0$. This is understood as an anharmonic vibration of the nuclear shape from a near spherical to a myriad of axial deformations encompassed by the second peak. The anharmonic behaviour of the Mo isotopes in this region was also reported in Ref.\cite{Kotila}.

A low lying excited $0^{+}$ state can be interpreted as the "ground state" for a shape isomer which has very distinct deformation. Such a structure is predominantly found in near-vibrational nuclei because the $U(5)$ model has the largest divergence of the equilibrium $\beta$ deformation between ground state and the first $\beta$ excited $0^{+}$ state, $(\langle\beta\rangle_{10}-\langle\beta\rangle_{00})/\langle\beta\rangle_{00}=0.3$. Besides having a lower $0^{+}$ state, the present model has also an even larger separation between the associated deformations, $(\langle\beta'\rangle_{10}-\langle\beta'\rangle_{00})/\langle\beta'\rangle_{00}=0.5$. This is another feature which supports its suitability for shape coexistence description at least within the considered nuclei.

The higher angular momentum experimental yrast sta\-tes available for $^{72}$Se and $^{72}$Kr are found to deviate from the vibrational level sequence. This is ascribed to the evolution of the higher states to a more pure prolate character which enhances the rotational motion in these nuclei \cite{Bender,Hinohara1}. Judging by the comparison made in Table \ref{tabe} between their normalized experimental energy spectrum and the theoretical one, the nuclei $^{72}$Se and $^{98}$Mo are found to be the best experimental realizations of the AEDC model. The experimental counterpart of the theoretical states is chosen to maximally match the theoretical results and which are without uncertainties in the angular momentum and parity assignment. Additionally, for levels with the same angular momentum, one associated the smallest vibrational quantum number $n$ to the lowest energy state. This is inspired by other $\gamma$-unstable solutions, such as for example $E(5)$ \cite{E5}, where the theoretical energy levels from distinct vibrational bands are no longer degenerate. The realization of the model in $^{72}$Se and $^{98}$Mo nuclei is also supported by the similar analysis in respect to the experimentally available $E2$ transition probabilities made in Table \ref{tabt}, where even if the experimental values are sizably overestimated by the theoretical predictions, those corresponding to the two mentioned nuclei are between the highest. A very good agreement with experiment is found for the transition $4_{0,2}\rightarrow2_{0,1}$ of $^{72}$Kr. This aspect, corroborated with a very good theoretical reproduction of the few available experimental low lying energy levels, makes this nucleus another suitable candidate for the AEDC. The experimental inband transitions of the considered nuclei are closer to the $U(5)$ values, whereas the interband ones are consistent with the decoupling predicted by the present formalism. While the $^{72}$Se nucleus was one of the first candidates for the shape coexistence \cite{Hamilton}, the $^{98}$Mo is a most recent addition \cite{Zielinska}. In consequence, the structure of the shape coexistence in $^{72}$Se is well established as having near spherical and prolate components \cite{Mc,Lju}. A similar interpretation is also proposed for $^{98}$Mo \cite{Thomas1,Thomas2,Zhang} with a mixture between spherical and $\gamma$-soft equilibrium shapes. Alternatively, a description through coexistence of various triaxial shapes \cite{Zielinska} was also offered for this nucleus. Therefore, the nature of the shape coexistence in this nucleus is far from being elucidated.

In general, it is possible to obtain an equivalent local potential associated to an energy dependent one \cite{Yekken1} having as input the exact energy spectrum. This is actually true only for singular potentials such as the Coulomb-like potential (\ref{pot}). In Ref.\cite{Yekken1} it was also shown that a Wood-Saxon-like potential is a quite good approximation for the energy dependent Coulomb potential. Transposing this information to the five-dimensional shape phase problem, one notes that the recently proposed Bohr-Mottelson model with Woods-Saxon potential \cite{Woods} fails to obtain a physically solid description of the $\beta$ band which is found to lay extremely low in energy. On the other hand, the corresponding equivalent potential in our case is more sharper in the origin, fact which allows a realistic description of the $\beta$ excited states which are still low due to the same finite structure of the outer barrier. Another advantage of the present formalism is that it reduces to an exactly solvable differential equation.

Although analytical characteristics of the AEDC are easy to grasp from the presented formulas and previous numerical analysis, its physical justification is not obvious. This is mostly because it is quite difficult to imagine an associated effective local potential, which otherwise holds the information regarding the physical behaviour of the system. However, one can draw some useful conclusions from the evolution of the system as function of $a$ in a convenient numerical interval of it. This is achieved in Fig.\ref{epot}, where for two different values of $a$ one plotted a set of states with their associated state-dependent potentials. The intersection between energy levels and potentials associated to the same state might be considered as points belonging to an effective local potential. Inspecting the two panels of Fig.\ref{epot}, one can see that as $a$ increases, the edge of the effective potential becomes sharper and the upper slope relevant for the energy spectrum decreases. This last aspect means that the system is softening when $a$ as well as the energy state increase. The rate of softening is obviously maximal in case of the AEDC. This interpretation is also supported by the behaviour of the $\beta$ density probability from Fig.\ref{ro}(a). In view of these arguments, AEDC seems to be a complementary model to the SSV. Indeed, while the latter corresponds to a fast stiffening nuclear surface, the present model describes an extremely $\beta$ soft nucleus.

Before closing this section it is necessary to comment about the apparent similarity of the energy dependent potential approach introduced in \cite{Eu} with the collective geometrical solutions obtained using a deformation dependent mass term \cite{M5,M6,M1,M2,M3}. The latter can be brought to a deformed Schr\"{o}dinger equation with an effective potential \cite{Quesne}. In the $\gamma$-unstable case of Kratzer potential \cite{M3}, which is an extension of the Coulomb potential, this effective potential contains additional dependence on seniority. The resemblance with the present approach stops at this point. First of all, one cannot construct analytically an effective local potential. This is because of the iterative nature of the associated differential equation containing explicitly its quantum number. Indeed, while the $\gamma$-unstable Coulomb-like potential leads to a similar $SO(2, 1)\times SO(5)$ \cite{FortVit} algebraic group structure as the Collective Geometric Model based on a Davidson potential \cite{Rowe1,Rowe2}, its energy dependent version cannot be expressed anymore in terms of the same generating operators because the energy dependent term must be replaced with the corresponding operator, tampering thus the involved commutation relations.

\section{Conclusions}

By considering a coupling constant for the $\gamma$-unstable Cou\-lomb\--like potential in the Bohr Hamiltonian which depends linearly on the system's energy, one obtained the corresponding energy spectrum and the wave functions as a function of the slope parameter. The analytical peculiarity of the induced energy dependence was duly investigated, imposing in the same time an existence interval for the slope parameter where the obtained formalism corresponds to a physically meaning system. As a result, the obtained energy spectrum is no longer bounded by the energy threshold corresponding to an infinite quantum number as in the unperturbed problem. Moreover the expansion of the energy spectrum was found to be saturated at normalized energy levels specific to vibrational states described by the $U(5)$ dynamical symmetry. Thus, in its asymptotic regime, the slope parameter acquires just a scaling role, providing in this way a new parameter free collective solution succinctly denoted AEDC, which along with other such models serves as reference points for general collective phenomena. Due to the specific structure of the model, the similarity with $U(5)$ stops at the energy level scheme. Indeed, as its analytical properties and the associated numerical applications show, the distribution of states by the seniority and $\beta$ vibration quantum numbers is different, with the special fingerprint of AEDC being the degeneracy of the first $0^{+}$ and $2^{+}$ states. Incidently this specific spectral signature is associated to shape coexistence phenomenon. The properties of the low lying states were also studied by means of the $\beta$ probability density distribution, which offered more insight into the distinct behaviour of AEDC as a model suitable for shape coexistence. Besides the downshift of the $\beta$ excited states in respect to the $U(5)$ ones, the $E2$ transition probabilities within AEDC are also different. The inband rates are overall higher, while the non-vanishing interband transitions are limited to $\Delta n=2$, such that consecutive $\beta$ excited bands are completely decoupled.

Experimental realization of the AEDC energy spectrum was found in the few nuclei $^{72}$Se, $^{74}$Se, $^{72}$Kr, $^{98}$Mo and $^{100}$Mo which are known to exhibit shape coexistence features. The presence of a low lying $0^{+}_{2}$ state was the  major criterium for candidates selection as it is the most pregnant spectral signature of the proposed model. The best representatives are considered $^{72}$Se and $^{98}$Mo nuclei. The experimental electromagnetic transitions of these isotopes however follow more closely the $U(5)$ predictions which hint to a strong spherical component underlying the shape coexistence in these nuclei. Even if the AEDC calculations are overestimated in all cases except $^{72}$Kr, the deviations are smaller for the two mentioned nuclei.

In conclusion, one must emphasize that this is the first fully collective attempt to describe the low lying energy spectrum of shape coexisting nuclei. The model obviously is capable to reproduce the experimental energy levels for these nuclei. However, for a more consistent description including the electromagnetic properties, the single-particle degrees of freedom are indispensable. There is however a way to extend the applicability of this model, by employing the Kratzer or Cornell potentials which are more pliable partners of the Coulomb potential.

\section*{Acknowledgments}
The author acknowledges the financial support received from the Romanian Ministry of Education and Research, through the Project PN-16-42-01-01/2016.

\section*{Appendix}

Various $\beta$ matrix elements needed for calculation of transition probabilities can be expressed analytically by using the following result \cite{Rass}:
\begin{eqnarray}
&&\int_{0}^{\infty}x^{\lambda}e^{-x}L_{n}^{s}(x)L_{n'}^{s'}(x)dx=(-)^{n+n'}\Gamma(\lambda+1)\nonumber\\
&\times&\sum_{k=0}^{Min(n,n')}{\lambda-s \choose n-k}{\lambda-s' \choose n'-k}{\lambda+k \choose k},\nonumber
\end{eqnarray}
which is valid for $Re(\lambda)>-1$. The factors defining the sum terms denote real extensions of the binomial coefficient.

%
% BibTeX users please use
% \bibliographystyle{}
% \bibliography{}
%
% Non-BibTeX users please use

\end{document}